\documentclass[a4paper,fleqn,usenatbib]{mnras}

\usepackage{newtxtext,newtxmath}

\usepackage[T1]{fontenc}
\usepackage{ae,aecompl}

\usepackage{graphicx}	
\usepackage{amsmath}	
\usepackage{amssymb}	

\usepackage[dvipsnames]{xcolor}
\usepackage{units}
\usepackage{textcomp}
\usepackage{colortbl}

\title[Modelling of mid-infrared interferometric signature of hot exozodiacal dust emission]{Modelling of mid-infrared interferometric signature of hot exozodiacal dust emission}

\author[F. Kirchschlager et al.
]{
Florian Kirchschlager$^{1,2}$\thanks{E-mail: f.kirchschlager@ucl.ac.uk}, Sebastian Wolf$^{2}$, Robert Brunngr\"aber$^{2}$, Alexis Matter$^{3}$,\newauthor Alexander V. Krivov$^{4}$, and  Aaron Labdon$^{5}$
\\
$^{1}$Department of Physics and Astronomy, University College London, Gower Street, London WC1E 6BT, United Kingdom\\
$^{2}$Institute of Theoretical Physics and Astrophysics, Kiel University, Leibnizstra\ss e 15, 24118 Kiel, Germany\\
$^{3}$ Laboratoire Lagrange, Universit\'e C\^ote d'Azur, Observatoire de la C\^ote d'Azur, CNRS, Boulevard de l'Observatoire, CS~34229,\\\phantom{$^{4}$ } 06304 Nice Cedex 4, France\\
$^{4}$Friedrich Schiller University Jena, Astrophysical Institute and University Observatory,  Schillerg\"a\ss chen 2-3, 07745 Jena, Germany\\
$^{5}$Department of Physics and Astronomy, University of Exeter, Stocker Road, Exeter EX4 4QL, United Kingdom\\
}

\date{Accepted XXX. Received YYY; in original form ZZZ}

\pubyear{2017}

\begin{document}
\label{firstpage}
\pagerange{\pageref{firstpage}--\pageref{lastpage}}
\maketitle

\begin{abstract}
Hot exozodiacal dust emission was detected in recent surveys around two dozen main sequence stars at distances of less than $\unit[1]{au}$ using H~and K~band interferometry. Due to the high contrast as well as the small angular distance between the circumstellar dust and the star, direct observation of this dust component is challenging. An alternative way to explore the hot exozodiacal dust is provided by mid-infrared interferometry. We analyze the L, M and N band interferometric signature of this emission in order to find stronger constraints for the properties and the origin of the hot exozodiacal dust. Considering the parameters of nine debris disc systems derived previously, we model the discs in each of these bands. We find that the M~band possesses the best conditions to detect hot dust emission, closely followed by L~and N~bands. The hot dust in three systems~-- HD~22484 (10~Tau), HD~102647 ($\beta$~Leo) and HD~177724 ($\zeta$~Aql)~--- shows a strong signal in the visibility functions which may even allow one to constrain the dust location. In particular, observations in the mid-infrared could help to determine whether the dust piles up at the sublimation radius or is located at radii up to $\unit[1]{au}$. In addition, we explore observations of the hot exozodiacal dust with the upcoming mid-infrared interferometer MATISSE at the Very Large Telescope Interferometer (VLTI).
\end{abstract}

\begin{keywords}
(stars:) circumstellar matter -- interplanetary medium -- infrared: planetary systems  -- techniques: interferometric
\end{keywords}
 

\section{Introduction}
\label{101}
\begin{table*} 
\caption{Parameters of targets with NIR excess observed with VLTI/VINCI (\citealt{Absil2009}) and CHARA/FLUOR (\citealt{Absil2013}) which is attributed to circumstellar dust.}
\centering
\label{data_systems}
 \hspace*{1cm}\begin{tabular}{r r c c c c c c c c c c} \hline 
HD\hspace*{0.2cm}& HIP\hspace*{0.2cm}& Alter.& $d$   &  $T_\star$&  $L_\star$ &Spectral     &Age& $R$& $M_\text{dust}$&$a_\text{max}$\\
number& number &name& [pc]                             &$\!$[K]              &[$\text{L}_\odot$]& class&[Gyr] &[au]&[$\unit[10^{-9}]{M_\oplus}$]&[$\text{\textmu m}$]\\\hline

10700&8102&$\tau$~Cet	        &$\phantom{1}3.7$       &5290&$\phantom{1}0.46$&G8$\,$V$\phantom{-V}$   &$10\phantom{.00}$& $0.016-0.056$&$0.03-\phantom{0}0.14$&$0.42$\\
22484&16852&10~Tau		&$13.7$          	&5998&$\phantom{1}3.06$&F9$\,$IV-V&$6.7$            & $0.039-0.49\phantom{0}$&$0.38-\phantom{0}6.95$&$0.71$\\
56537&35350&$\lambda$~Gem	&$28.9$       		&7932&$27.4\phantom{0}$&A3$\,$V$\phantom{-V}$   &$0.5$            & $0.12\phantom{0}-0.65\phantom{0}$&$0.69-\phantom{0}3.1\phantom{0}$&$0.42$\\
102647&57632&$\beta$~Leo	&$11.1$			&8604&$13.25$	    &A3$\,$V$\phantom{-V}$      &$0.1$            & $0.08\phantom{0}-0.98\phantom{0}$&$0.57-\phantom{0}5.7\phantom{0}$&$0.77$\\
172167&91262&$\alpha$~Lyr	&$\phantom{1}7.8$       &9620&$37\phantom{.70}$&A0$\,$V$\phantom{-V}$	  &$0.7$            & $0.15\phantom{0}-0.65\phantom{0}$&$0.89-\phantom{0}3.5\phantom{0}$&$0.35$\\
177724&93747&$\zeta$~Aql	&$25.5$                 &9078&$36.56$	    &A0$\,$IV-V	  &$0.8$            & $0.14\phantom{0}-1.0\phantom{00}$&$1.79-10.2\phantom{0}$&$0.51$\\
187642&97649&$\alpha$~Aql	&$\phantom{1}5.1$       &7680&$10.2\phantom{0}$&A7$\,$IV-V&$1.3$            & $0.09\phantom{0} -0.11\phantom{0}$&$0.84-\phantom{0}1.0\phantom{0}$& $0.2\phantom{0}$ \\
203280&105199&$\alpha$~Cep	&$15.0$                 &7700&$19.97$          &A7$\,$IV-V&$0.8$            & $0.11\phantom{0} -0.43\phantom{0}$&$0.54-\phantom{0}2.1\phantom{0}$&$0.35$\\ 
216956&113368&$\alpha$~PsA	&$\phantom{1}7.7$       &8590&$16.6\phantom{0}$&A3$\,$V$\phantom{-V}$	  &$0.4$            & $0.11\phantom{0} -0.17\phantom{0}$&$0.12-\phantom{0}0.2\phantom{0}$&$0.25$\\ 
\hline 
 \end{tabular}
\newline
\raggedright
\textbf{References:} The distances are derived from parallax measurements (\citealt{vanLeeuwen2007}). Stellar temperatures and luminosities are adopted from  \cite{Habing2001}, \cite{vanBelle2001}, \cite{vanBelle2006}, \cite{Mueller2010}, \cite{Zorec2012}, and \cite{Boyajian2013}, spectral classes and stellar ages from \cite{Mamajek2012}, \cite{Vican2012}, and \cite{Absil2013}, and disc and dust constraints are from \cite{Kirchschlager2017}.
\end{table*}
Hot exozodiacal dust is located at the closest possible distances to main sequence-stars (\citealt{Absil2006}). Due to its high temperature the dust is traced in observations in the near-infrared (e.$\,$g. \citealt{Akeson2009}; \citealt{Defrere2011}). According to the sparing amount of exozodiacal dust in circumstellar environments, the emitted fluxes are quite low, and the excesses above the stellar photospheres are marginal. Therefore, detection of circumstellar emission caused by \mbox{exo}zodiacal dust was so far only possible through interferometric observations (\citealt{Absil2013}; \citealt{Ertel2014b}; \citealt{Nunez2017}). 

Based on near- and mid-infrared (NIR; MIR) observations of a sample of nine debris disc systems harbouring hot exozodiacal dust, \cite{Kirchschlager2017} constrained the dust properties and dust distribution in the vicinity of these stars. The dust was found to be located within $\unit[\sim0.01-1]{au}$, and the exozodiacal dust masses amounted to $\unit[(0.2{-}3.5)\times10^{-9}]{M_\oplus}$. The dust grains were determined to be below $\unit[0.2-0.5]{\text{\textmu}\textrm{m}}$.

The origin of the hot exozodiacal dust is still highly debated and explaining the presence of small dust grains located close to the star in sufficient large amount over a long time is challenging. Proposed scenarios are the aftermath of a large collision or heavy bombardement, dynamically perturbed comets, planetesimals or planets, which move and disintegrate in the inner system,  the sublimation of a supercomet close to the star, an inward drift of dust grains caused by the Poynting-Robertson drag, as well as charged nanoscale dust grains trapped in a stellar magnetic field (\citealt{Kral2017}, and references therein). However, each attempt to explain the presence of hot exozodiacal dust suffers from inconsistencies. Therefore, further observations are necessary to shed light on the properties of the hot dust.

While photometry of the dust emission at wavelengths below $\unit[15]{\text{\textmu}\textrm{m}}$ suffers from the uncertainty due to small fluxes compared to the stellar spectrum, interferometric observations provide a direct estimation of the flux ratio between the circumstellar environment and the stellar photosphere. Well determined  NIR and MIR fluxes are necessary to place tighter constraints on the parameters of the hot exozodiacal dust. Therefore, complementary interferometric observations in this wavelength range are required.

MATISSE (Multi AperTure mid-Infrared SpectroScopic Experiment; \citealt{Lopez2014}) is an upcoming second generation VLTI instrument which will offer simultaneous four-beam interferometry and spectroscopic capabilities in L ($\lambda=\unit[2.8-4.0]{\text{\textmu}\textrm{m}}$), M ($\unit[4.5-5.0]{\text{\textmu}\textrm{m}}$) and N band ($\unit[8-13]{\text{\textmu}\textrm{m}}$). Especially, the new possibility of interferometric observations in L and M band will improve and complete the investigation of milliarcsecond scale MIR emission (\citealt{Matter2016a}).

In this paper, we investigate the mid-infrared interferometric signature of hot exozodiacal dust emission around nine main-sequence stars. For this purpose, we use model parameters derived by \cite{Kirchschlager2017}, calculate thermal reemission and scattered light maps and derive the corresponding visibility functions. The applied disc and dust models are described in Section~\ref{201} and the results are presented in Section~\ref{301}. In Section~\ref{401} we summarize our findings and have an outlook on observations of hot exozodiacal dust with the upcoming instrument MATISSE.


\section{Stellar sample and model}
\label{201}
In this Section, we present the strategy to model and evaluate the interferometric signature of the exozodiacal dust belt in the nine selected systems. The investigated sample of systems is introduced in Section~\ref{sec_survey} and the applied model is described in Section~\ref{sec_model}. In Section~\ref{sec_visib} the procedure to simulate visibility functions of these systems is given.

\subsection{Stellar sample}
\label{sec_survey}
In Table~\ref{data_systems}, those debris disc systems are compiled which show a significant NIR~excess in K~band ($\unit[1.59 - 1.77]{\text{\textmu}\textrm{m}}$) that is indicative of hot exozodiacal dust (VLTI/VINCI, \citealt{Absil2009}; CHARA/FLUOR, \citealt{Absil2013}). In addition, interferometric flux measurements at \mbox{$\lambda\approx\unit[10]{\text{\textmu}\textrm{m}}$} (\citealt{Mennesson2014}) exist for these systems, which strongly constrain the disc and dust parameters (\citealt{Kirchschlager2017}). All of these stars are on or close to the main sequence, with spectral classes ranging from G8 to A0 and luminosities from $L_\star\sim\unit[0.4]{L_\odot}$ up to $\unit[\sim40]{L_\odot}$.  The stellar fluxes are in the range of $\unit[20-263]{Jy}$ and $\unit[3-39]{Jy}$ for the L and N~band, respectively.

\subsection{Model description}
\label{sec_model}
For the sake of conformity, we use the same simple model as in \cite{Kirchschlager2017}.

{\it Disc properties:} 
The disc model consists of a thin, radially symmetric ring with an inner radius $R$ and outer radius of $1.5\,R$. The number density decreases with distance as $n(r)\propto r^{-1}$. The disc is assumed to be face-on ($i=0^\circ$) and the half opening angle amounts to $5^\circ$. 

{\it Dust properties:} Each dust ring is composed of compact, spherical dust grains with grain radius $a$. The grains are assumed to consist of pure graphite ($\rho=2.24\,$g\,cm$^{-3}$; \citealt{WeingartnerDraine2001}), using the 1/3$-$2/3 approximation (\citealt{DraineMalhotra1993}). Silicate compositions could be excluded by the modelling of \cite{Kirchschlager2017}, since the presence of the prominent silicate feature at $\unit[\sim10.0]{\text{\textmu}\textrm{m}}$ would contradict the observed MIR fluxes.

\begin{figure*}
\centering
\includegraphics[trim=2.1cm 12.4cm 2.4cm 12.01cm, clip=true,page=1,width=0.99\linewidth]{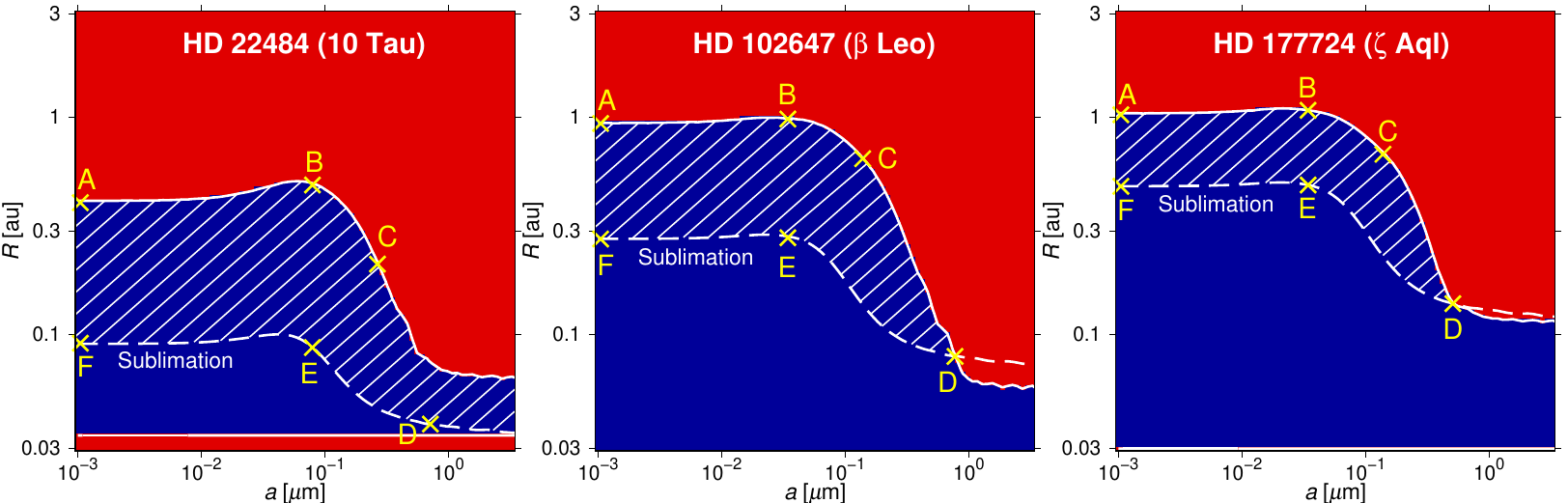}
\vspace*{-0.1cm} 
\caption{Parameter space explored for the systems HD~22484, HD~102647 and HD~177724 (\citealt{Kirchschlager2017}).  Blue regions correspond to models which possess simulated NIR and MIR fluxes which are consistent to previous observations within their errors, red regions fail. The shaded areas correspond to parameter settings which allow the existence of hot exozodiacal dust, and which is constrained by observational data (blue-red border, white solid line) and the sublimation radius (white dashed line). The configurations \mbox{A -- F} are the extremes of the model space  selected for the visibility calculations.}
\label{Para1a} 
\end{figure*}

\cite{Kirchschlager2017} showed that the grain radii of the hot exozodiacal dust in most of the systems must be below $\unit[0.5]{\text{\textmu}\textrm{m}}$. However, in the case of HD~10700 ($\tau$~Cet), HD~22484 (10~Tau) and HD~56537 ($\lambda$~Gem), the presence of micrometer-sized grains cannot be ruled out, 
since only MIR~observations (\citealt{Mennesson2014}) with a limited resolution are available for these systems. Nevertheless, we expect that micrometer-sized grains are absent in these systems, too. We estimate the maximum grain size in these three systems by extrapolating the steepest part of the blue-red border, around point C, to larger sizes (see Fig.~\ref{Para1a} for the system HD~22484). 

As shown in Figure~\ref{Para1a} (adopted from \citealt{Kirchschlager2017}), there exists an area in the parameter space ($a$, $R$) in which the models can reproduce previous observations\footnote{In this area of the parameter space, the models are consistent with the observational data within their error bars (\citealt{Kirchschlager2017}).} in the NIR (VLTI/VINCI, FLUOR/CHARA; \citealt{Absil2009,Absil2013}) and MIR (Keck; \citealt{Mennesson2014}) while the dust is still not sublimated. Details for the calculation of the sublimation radius are given in \cite{Kirchschlager2017}. While the analysis of the entire parameter space would result in extensive data, we instead  explore our simulations for each system at six representative parameter configurations \mbox{(A -- F)} at the various borders of the possible parameter range, enabling to study the impact of individual parameters on the visibility functions with an appropriate computational effort. The corresponding grain size $a$, disc ring radius $R$ and the dust mass $M_\text{dust}$ of each configuration are used for the modelling of the individual systems. The configurations A, B, and C correspond to larger disc extensions and higher dust masses than the configurations D, E, and F which are defined by the sublimation radius. The grain size increases in the alphabetical order from configuration A to D, and then decreases to configuration~F.  In Table~\ref{ParaHD102647}, we list the parameter settings for the systems HD~22484, HD~102647 and HD~177724. 

\subsection{Procedure and software applied}
\label{sec_visib}
\begin{table} 
\caption{Parameters of the configurations A -- F corresponding to the acceptable parameter space (Fig.~\ref{Para1a}) for the systems HD~22484, HD~102647 and HD~177724.}
\label{ParaHD102647}
\centering
 \begin{tabular}{c l l l l  l l l}\hline 
\multicolumn{2}{c}{\bf HD~22484}&$\hspace*{-0.3cm}\phantom{0}$A&$\phantom{0}$B&$\phantom{0}$C&$\phantom{0}$D&$\phantom{0}$E&$\phantom{0}$F\\\hline
$\hspace*{-0.10cm}R$&$\hspace*{-0.3cm}$[au]					&$\hspace*{-0.10cm}0.41$&$\hspace*{-0.10cm}0.49$&$\hspace*{-0.10cm}0.21$&$\hspace*{-0.10cm}0.04$&$\hspace*{-0.10cm}0.09$&$\hspace*{-0.10cm}0.09$\\
$\hspace*{-0.10cm}a$&$\hspace*{-0.3cm}$[$\text{\textmu m}$]                       &$\hspace*{-0.10cm}0.001$&$\hspace*{-0.10cm}0.079$&$\hspace*{-0.10cm}0.266$&$\hspace*{-0.10cm}0.712$&$\hspace*{-0.10cm}0.079$&$\hspace*{-0.10cm}0.001$\\
$\hspace*{-0.10cm}M_\text{dust}$&$\hspace*{-0.3cm}$[$\unit[10^{-9}]{M_\oplus}$]    &$\hspace*{-0.10cm}6.94$&$\hspace*{-0.10cm}6.72$&$\hspace*{-0.10cm}3.62$&$\hspace*{-0.10cm}1.23$&$\hspace*{-0.10cm}0.38$&$\hspace*{-0.10cm}0.53$\\
\hline 
 \end{tabular}
 \begin{tabular}{c l l l l  l l l}\hline 
\multicolumn{2}{c}{\bf HD~102647}&$\hspace*{-0.3cm}\phantom{0}$A&$\phantom{0}$B&$\phantom{0}$C&$\phantom{0}$D&$\phantom{0}$E&$\phantom{0}$F\\\hline
$\hspace*{-0.10cm}R$&$\hspace*{-0.3cm}$[au]					&$\hspace*{-0.10cm}0.94$&$\hspace*{-0.10cm}0.98$&$\hspace*{-0.10cm}0.65$&$\hspace*{-0.10cm}0.08$&$\hspace*{-0.10cm}0.28$&$\hspace*{-0.10cm}0.28$\\
$\hspace*{-0.10cm}a$&$\hspace*{-0.3cm}$[$\text{\textmu m}$]                       &$\hspace*{-0.10cm}0.001$&$\hspace*{-0.10cm}0.034$&$\hspace*{-0.10cm}0.138$&$\hspace*{-0.10cm}0.77$&$\hspace*{-0.10cm}0.034$&$\hspace*{-0.10cm}0.001$\\
$\hspace*{-0.10cm}M_\text{dust}$&$\hspace*{-0.3cm}$[$\unit[10^{-9}]{M_\oplus}$]    &$\hspace*{-0.10cm}5.73$&$\hspace*{-0.10cm}5.66$&$\hspace*{-0.10cm}5.17$&$\hspace*{-0.10cm}0.57$&$\hspace*{-0.10cm}0.7$&$\hspace*{-0.10cm}0.75$\\
\hline 
 \end{tabular}
 \begin{tabular}{c l l l l  l l l}\hline 
\multicolumn{2}{c}{\bf HD~177724}&$\hspace*{-0.3cm}\phantom{0}$A&$\phantom{0}$B&$\phantom{0}$C&$\phantom{0}$D&$\phantom{0}$E&$\phantom{0}$F\\\hline
$\hspace*{-0.10cm}R$&$\hspace*{-0.3cm}$[au]					&$\hspace*{-0.10cm}1.03$&$\hspace*{-0.10cm}1.08$&$\hspace*{-0.10cm}0.68$&$\hspace*{-0.10cm}0.14$&$\hspace*{-0.10cm}0.49$&$\hspace*{-0.10cm}0.48$\\
$\hspace*{-0.10cm}a$&$\hspace*{-0.3cm}$[$\text{\textmu m}$]                       &$\hspace*{-0.10cm}0.001$&$\hspace*{-0.10cm}0.034$&$\hspace*{-0.10cm}0.138$&$\hspace*{-0.10cm}0.509$&$\hspace*{-0.10cm}0.034$&$\hspace*{-0.10cm}0.001$\\
$\hspace*{-0.10cm}M_\text{dust}$&$\hspace*{-0.3cm}$[$\unit[10^{-9}]{M_\oplus}$]    &$\hspace*{-0.10cm}10.1$&$\hspace*{-0.10cm}10.2$&$\hspace*{-0.10cm}9.04$&$\hspace*{-0.10cm}1.79$&$\hspace*{-0.10cm}3.09$&$\hspace*{-0.10cm}3.23$\\
\hline 
 \end{tabular}
\end{table}
The optical properties of the grains are calculated with the software tool \textsc{miex} which is based on the Mie scattering theory (\citealt{Mie1908}; \citealt{WolfVoshchinnikov04}). For each parameter configuration (A -- F), single scattering and reemission simulations are performed and the maps are calculated using an enhanced version of the tool \textsc{debris} (\citealt{Ertel2011}). The simulations are performed in the L ($\lambda=\unit[3.5]{\text{\textmu}\textrm{m}}$), M ($\unit[4.7]{\text{\textmu}\textrm{m}}$), and N band ($\unit[10.0]{\text{\textmu}\textrm{m}}$).

Subsequently, the brightness distributions are integrated along one direction in the disc and the visibility functions are derived along the perpendicular direction. Using the Van Cittert-Zernike theorem, the intensity map is used to determine the visibility of the system via Fourier transformation as a function of projected telescope baseline length. 

In addition, for each of the nine systems we calculate the visibility function $V_\star$ of the single star only. We determine the difference $\Delta V = V_\star-V_{\text{(}\star \text{ + disc)}}$ of the visibility functions of the single star and the one of the system consisting of star and disc at baseline lengths of $\unit[20-140]{m}$. When the difference $\Delta V$ is significant, the visibilities $V_{\text{(}\star\text{ + disc)}}$ and $V_{\star}$ are directly connected to the flux ratio $f$ between disc and central star via $V_{\text{(}\star\text{ + disc)}} \approx(1-f)\text{V}_{\star} $ (\citealt{DiFolco2004}). Therefore, the flux emitted by the dust in the wavebands can be derived by the detection of a significant visibility deficit. 

\section{Results}
\label{301}

In this section the results of the calculated visibility functions in the L ($\lambda=\unit[3.5]{\text{\textmu}\textrm{m}}$), M ($\unit[4.7]{\text{\textmu}\textrm{m}}$), and N band ($\unit[10.0]{\text{\textmu}\textrm{m}}$) are presented. We will see that HD~22484, HD~102647 and HD~177724 are the most promising systems.


\subsection{Visibility drop}
In five of nine systems of Table~\ref{data_systems} the visibility difference amounts to values larger than or equal to $\unit[2]{per~cent}$ (Tab.~\ref{results_systems}), namely HD~22484, HD~102647, HD~172167, HD~177724 and HD~187642. The results of the configurations A, B and C, which correspond to larger disc extensions and higher dust masses, show distinctly higher visibility differences than in the cases for the configurations E and F, which correspond to dust locations close to the sublimation radius. Except for configuration D in HD~22484, the visibility differences are up to $\unit[5]{per~cent}$ in the L~and M~band and $\unit[4]{per~cent}$ in the N~band. In general, the visibility differences for the L and N~bands are slightly smaller than the ones for the M~band. For configuration D in HD~22484, the differences are up to $\unit[{\sim}10]{per~cent}$ for both the L and M~band and $\unit[3]{per~cent}$ in the N~band.

\begin{table}
\caption{Visibility differences $\Delta V$ for the simulated interferometric observations in the M~band and at baseline lengths in the range $\unit[20-140]{m}$. The values for the most promising systems HD~22484, HD~102647 and HD~177724 are highlighted. In general, the values for the L and N~bands are only slightly smaller.}
\centering
\label{results_systems}
\begin{tabular}{r c c} \hline 
HD\hspace*{0.2cm}&\multicolumn{2}{c}{$\Delta V$ at configurations}\\
number&A,$\,$B,$\,$C$\,[$per~cent$]$  &D,$\,$E,$\,$F$\,[$per~cent$]$  \\\hline
10700&$0.7-1.3$ &$0.2-1.7$ \\
\rowcolor{lightgray}22484&$2.5-4.5$&$0.5-{\sim}10$ \\
56537&$0.7-1.4$ &$0.2-1.6$ \\
\rowcolor{lightgray}102647&$2.5-4\phantom{.0}$ &$0.5-3\phantom{.0}$ \\
172167&$1.5-2\phantom{.0}$ &$1.2-2.5$ \\
\rowcolor{lightgray}177724&$2\phantom{.0}-3.5$ &$2\phantom{.0}-4.5$ \\
187642&$1.5-3.5$ &$1.5-3.5$ \\
203280&$0.8-1.3$ &$0.4-1.6$ \\
216956&$0.2-0.4$ &$0.2-0.8$ \\ 
\hline 
 \end{tabular}
\end{table}

  \begin{figure} 
 \includegraphics[trim=1.95cm 0.25cm 2.9cm 0.1cm, clip=true,width=1.0\linewidth, page=1]{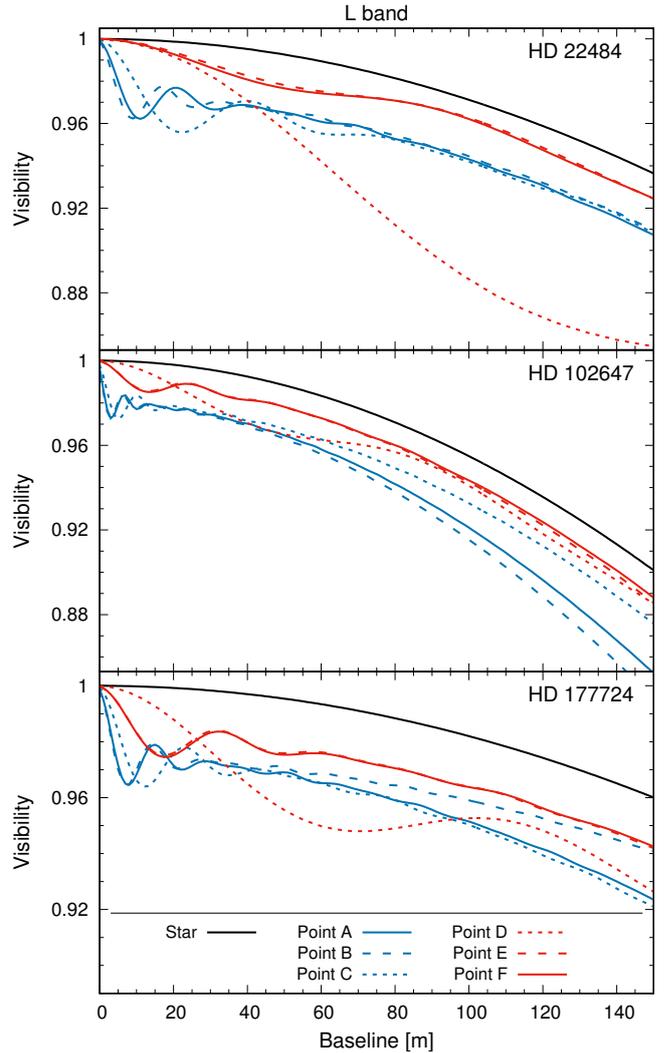}
\vspace*{-0.4cm}
\caption{L~band visibilities  ($\lambda=\unit[3.5]{\text{\textmu}\textrm{m}}$) for three systems harbouring exozodiacal dust: HD~22484 (top), HD~102647 (middle) and HD~177724 (bottom). The black solid line indicates the visibility of the central star without disc ($V_\star$). Different configurations in the parameter space (see Fig.~\ref{Para1a}) are indicated by different colours and line types.}
\label{Results_three_systems_visib_L}
 \end{figure} 

 \begin{figure} 
 \includegraphics[trim=1.95cm 0.25cm 2.9cm 0.1cm, clip=true,width=1.0\linewidth, page=1]{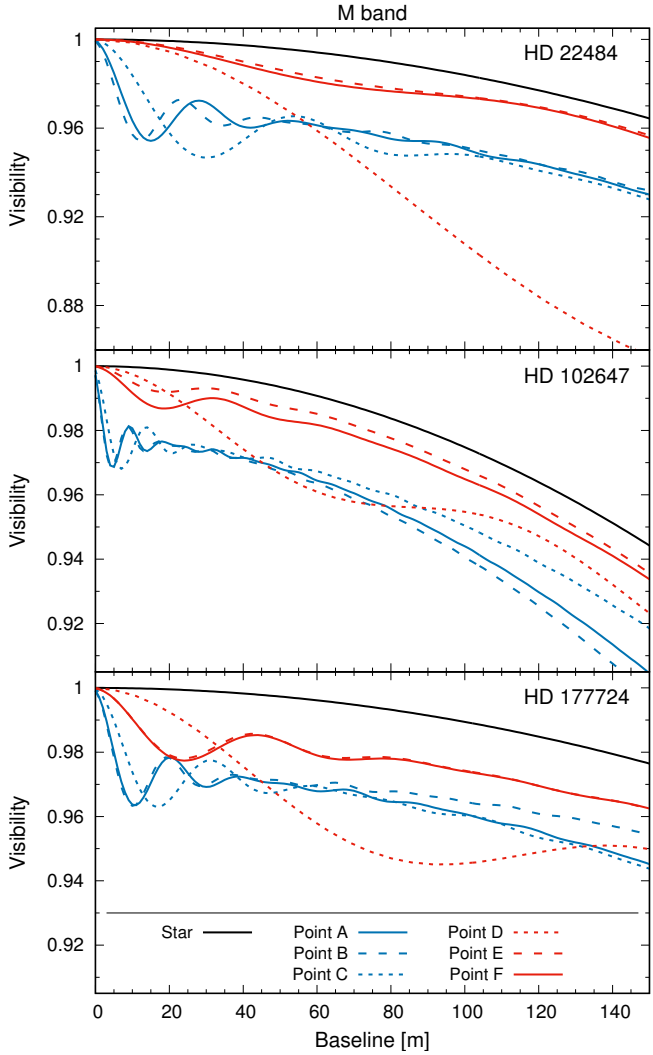} 
\vspace*{-0.4cm}
\caption{Same as Fig.~\ref{Results_three_systems_visib_L}, but for the M~band ($\lambda=\unit[4.7]{\text{\textmu}\textrm{m}}$).}
\label{Results_three_systems_visib_M}
 \end{figure} 

 \begin{figure} 
 \includegraphics[trim=1.95cm 0.25cm 2.9cm 0.1cm, clip=true,width=1.0\linewidth, page=1]{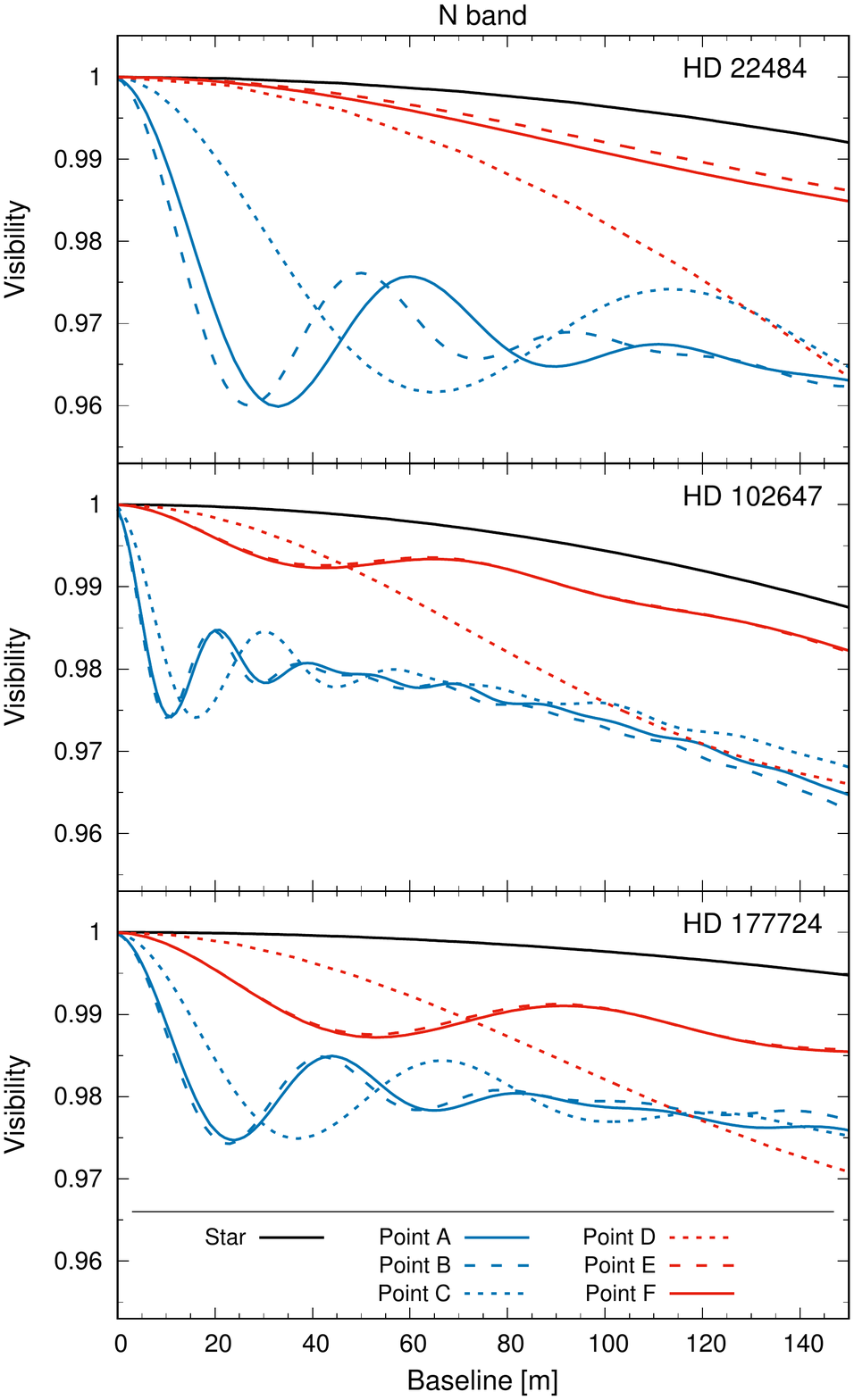} 
\vspace*{-0.4cm}
\caption{Same as Fig.~\ref{Results_three_systems_visib_L} and \ref{Results_three_systems_visib_M}, but for the N~band ($\lambda=\unit[10]{\text{\textmu}\textrm{m}}$).}
\label{Results_three_systems_visib_N}
 \end{figure} 

The hot exozodiacal dust emission in the system of HD~216956 amounts to less than $\unit[0.8]{per~cent}$ and hence constitutes the system with the lowest visibility difference. For the remaining three systems -- HD~10700, HD~56537 and HD~203280 -- the hot dust emission causes a visibility drop lower than  $\unit[1.7]{per~cent}$.

Overall, these results show that a larger disc ring radius $R$ produces a larger visibility deficit, while there is a weaker impact of the dust grain size.

 \subsection{Constraints for the dust properties}
Observations of the systems HD~22484, HD~102647, and HD~177724 potentially allow one to derive constraints for the dust location (Figs.~\ref{Results_three_systems_visib_L}--\ref{Results_three_systems_visib_N}). For these systems, the L~and M~band visibility differences between the blue and red curves are larger than or in the order of $\unit[1.5]{per~cent}$. Using an instrument with a 1$\sigma$ visibility accuracy of $\unit[{\sim}0.5]{per~cent}$, L~and M~band observations provide the possibility to significantly (3$\sigma$) distinguish  between the configurations A, B, C (larger disc radius) and the configurations E, F (sublimation radius). In other words, one could distinguish between emission from dust grains piled up at the sublimation radius and emission from dust grains located at a ten times larger radial distance.
However,  configuration D hampers a clear allocation of the location of the dust emission, as the visibility curve of configuration D intersects the blue curves in all three systems in both L~and M~band. For HD~22484 and HD~102647, N~band observations at baselines smaller than $\unit[100]{m}$ and $\unit[65]{m}$, respectively, are able to avoid a confusion with configuration D. For HD~177724, however, configuration D can not be distinguished clearly to configurations which correspond to larger discs extensions. A non-detection of dust emission at baselines longer than $\unit[100]{m}$ would indicate dust located near to the sublimation radius and grain sizes smaller than $\unit[0.7]{\text{\textmu}\textrm{m}}$ in these three systems.

Overall, MIR interferometric observations potentially allow one to constrain the parameters of the hot exozodiacal dust if the 1$\sigma$ visibility accuracy of the instrument amounts to $\unit[{\sim}0.5]{per~cent}$.

These constraints might help to reveal the origin of the dust material, which is still highly debated. \cite{Kirchschlager2017} suggested the possibility that the grain temperatures of the exozodiacal dust belts are the same in all systems. We checked that MIR interferometric observations would not be able to confirm or exclude this hypothesis. The possible temperature ranges are too wide to make any further restrictions.

\section{Conclusions and outlook}
\label{401}
We conclude our study with a summary of the results of MIR interferometric observations of the hot exozodiacal dust belts found in Section~\ref{301}. Based on constraints of the disc and dust distributions derived in \cite{Kirchschlager2017}, we investigated the possibility to detect hot exozodiacal dust emission around nine main-sequence stars using MIR interferometry. Our conclusions are as follows:
\begin{itemize}
\item The visibility drop caused by the hot exozodiacal dust is in five of the nine systems of Table~\ref{data_systems} -- HD~22484, HD~102647, HD~172167, HD~177724 and HD~187642 -- larger than $\unit[2]{per~cent}$. In general, the visibility differences for the L and N~band are slightly smaller than the ones for the M~band.
\item MIR interferometric observations will potentially allow one to constrain the parameters of the hot exozodiacal dust systems and hence might help to reveal the origin of the dust material. Assuming a 1$\sigma$ visibility accuracy of $\unit[{\sim}0.5]{per~cent}$, a combination of L,~M~and N~band observations has the potential to distinguish between emission of dust grains at the sublimation radius and emission of dust grains located at larger distances. The most promising systems are HD~22484 and HD~102647, followed by HD~177724. Both the detection and the non-detection may help to constrain the dust location.
\end{itemize}
As an outlook, the upcoming instrument \mbox{MATISSE} at the VLTI will specifically perform simultaneous observations in the L,~M~and N~band, which would make it a very suitable instrument for the characterization of the hot exozodiacal dust. However, such a study appears challenging for \mbox{MATISSE} given the required visibility accuracy of less than $\unit[1]{per~cent}$. Indeed, theoretical \mbox{MATISSE} visibility accuracies of 1 to $\unit[3]{per~cent}$ in L~and M~band, and $\unit[8]{per~cent}$ in N~band, were estimated for a $\unit[20]{Jy}$ source; estimates were based on SNR calculations including the contribution of the fundamental noises (source photon noise, readout noise, thermal background photon noise) and the transfer function variations \citep{Lopez2014,Matter2016b}. Nevertheless, in the frame of the recent \mbox{MATISSE} test phase, several lab transfer function measurements were performed over a few hours with a very bright artificial IR source. Such a source would have an equivalent flux, if observed with the UTs, of 20 to $\unit[70]{Jy}$ in N-band, $\unit[400]{Jy}$ in M-band, and $\unit[600]{Jy}$ in L~band. This lead to absolute visibility accuracies of $\unit[0.5]{per~cent}$ in L~band, $\unit[0.4]{per~cent}$ in M~band, and less than $\unit[2.5]{per~cent}$ in N~band, on average over the corresponding spectral band. Those promising results are extensively described in internal ESO documents written by the \mbox{MATISSE} consortium (private communication with A. Matter). Eventually, a proper study of the detectability of exozodiacal dust with \mbox{MATISSE} will require a knowledge of its real on-sky performance, which will notably include the effects of the sky thermal background fluctuation, the atmospheric turbulence, or the on-sky calibration. The beginning of the on-sky tests (commissioning) of \mbox{MATISSE} is planned for March 2018.

\section*{Acknowledgements}
F.K., S.W., R.B., and A.V.K. thank the DFG for financial support under contracts WO 857/13-1, WO 857/15-1, and KR 2164/15-1.
 


  \bibliographystyle{mnras}
{\footnotesize
  \bibliography{Literature}
}
\label{lastpage}

\bsp	

\end{document}